\documentstyle[aps,epsfig,float,twocolumn,prl]{revtex} 
\newcommand{\be}{\begin{equation}}
\newcommand{\ee}{\end{equation}}

\begin{document}
\twocolumn[\hsize\textwidth\columnwidth\hsize\csname @twocolumnfalse\endcsname 
\draft
\title{Do Vortices Entangle?}
\author{C.J. Olson Reichhardt and M. B. Hastings}
\address{
Center for Nonlinear Studies and Theoretical Division, Los Alamos National
Laboratory, Los Alamos, New Mexico 87545
}
\date{\today}
\maketitle
\begin{abstract}
We propose an experiment for directly constructing and locally probing
topologically entangled states of superconducting vortices
which can be performed with present-day
technology.  
Calculations using an elastic string vortex model indicate
that as the pitch (the winding angle divided by the vertical
distance) increases, the vortices approach each other.
At values of the pitch higher than a maximum value  
the entangled
state becomes unstable to collapse via a singularity of the model.  
We provide predicted experimental signatures for both vortex
entanglement and vortex cutting.  The local probe we propose can also
be used to explore a wide range of other quantities.
\vskip2mm
\end{abstract}
\pacs{PACS: 74.25.Qt, 74.25.Sv}
]

The high superconducting transition temperatures of 
compounds such as YBa$_2$Cu$_3$O$_{(7-\delta)}$ (YBCO) and 
BiSr$_2$Ca$_2$CuO$_8$ (BSCCO) 
lead to a very rich set of behaviors
of the 
magnetic vortices which
form inside the material in the presence of an applied magnetic
field. Thermal fluctuations are significant over a wide range of the
(H,T) phase diagram \cite{Blatter94}, 
causing the lattice of stiff vortex lines to
melt well below $T_c$.  The nature of this molten state has remained
a subject of intense debate.  

Since there can be significant 
thermally-induced wiggling along the length of the vortex in the
liquid state, Nelson proposed that neighboring vortex lines may 
become entangled with each other, in analogy with a superfluidity
transition in a boson system \cite{Nelson88}.
This entanglement could produce a dramatic increase in the 
viscosity of the vortex liquid \cite{Nelson89,Marchetti90,Obukhov90},
similar to that which occurs for entangled polymers.
As a result, vortex pinning by random disorder in the sample
would be enhanced, so that the resistivity
would drop in the entangled state.

In order for the vortices to entangle,
it is crucial that neighboring lines not cut through
each other easily \cite{Clem82} and reconnect into a disentangled
state.  Estimates of the cutting barrier vary widely 
\cite{Nelson89,Obukhov90,Marchetti91b,Sudbo91,CutBar,Carraro95}, 
ranging from $50k_BT$ to of order $k_BT$, leaving the question of 
whether vortices can easily cut in the liquid state unresolved.
Numerical simulations performed in limits ranging from the low-field London
regime to the high-field lowest Landau level regime have proven
similarly ambiguous, with some simulations interpreted as providing
evidence for entanglement \cite{EntangSim}
and others interpreted as showing that the lines cut and do not
entangle
\cite{CutBar,Li94,CutSim}.
The simulations are limited both by the models chosen and by
the system sizes that can be simulated.
Models based
on the boson analogy lack long-range interactions along the vortex
line, which can stiffen the vortices and might reduce 
entanglement.
\cite{Marchetti91b,Sudbo91b,Benetatos9900}.
In the frustrated 3D XY model, there are multiple
ways to define a path followed by a vortex line, some of which are consistent
with entanglement, and others which are not \cite{Li94}.

Since theoretical and numerical evidence for entanglement have proven
inconclusive up to this point, it is natural to turn to experiments to
resolve the issue.
Unfortunately experimental evidence for or against entanglement 
\cite{Lopez98Puig00,Qiu98,Miu98,Grigera02}
has also proven ambiguous, both because the experimental measures are indirect,
and because there is considerable variation in the definition of entanglement
\cite{Nelson95Ertas96},
such as equating entanglement with the onset of plasticity 
\cite{Lopez98Puig00}.
Thus, despite more than a decade of theoretical, numerical, and experimental
studies, the question of whether vortices in high-temperature superconductors
can form an entangled state has not yet been convincingly answered.

Here we propose a direct experimental test of vortex entanglement
by means of a local atomic-force microscope (AFM) 
probe which can unambiguously determine whether
it is possible for two vortices to wind around each other without cutting.
We consider conditions that are as favorable as possible for
entanglement: low vortex density, weak background pinning, and low
temperatures, which should increase the barrier to flux cutting.
We propose using a magnetic AFM tip 
to produce the simplest possible entangled state,
two vortices wound around each other, and show theoretically that
the forces involved in creating the entangled state
can be measured with current technology.
Should the vortices cut rather than entangling, a distinct force
signature will be observed.
If entanglement does not occur in this limit, it seems unlikely to
occur in the higher density, higher temperature conditions near
the melting transition.  In addition to testing entanglement, the
local AFM probe which we propose can be used to measure a wide range of
other quantities, such as vortex line tension, local pinning force,
and local shear forces.

We use as our definition of an entangled vortex state the picture
originally proposed by Nelson, in which vortices behave like elastic
lines that cannot easily cut through each other \cite{Nelson88}.
We consider a layered superconductor containing two closely spaced magnetic
dots at the bottom of the sample \cite{Schuller} (dark circles in Fig. 1) 
A mobile magnetic dot
is introduced to the top of the sample 
in the form of a magnetic AFM 
tip (open circle in Fig. 1).
An optional additional magnetic dot 

\begin{figure}
\begin{center}
\epsfxsize=3.5in
\epsfbox{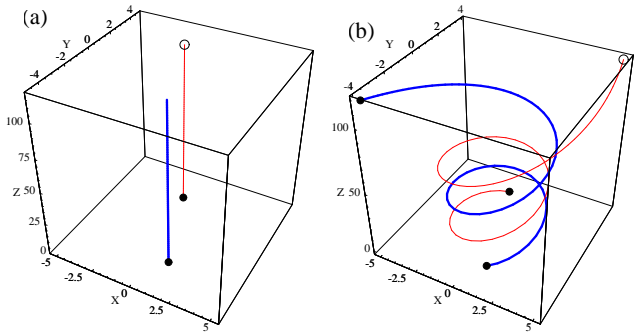}
\end{center}
\caption{(a) Schematic of the starting state. Black dots 
represent the fixed pins; the open circle 
represents the end of the vortex being moved by the AFM.
(b)
A configuration of wound vortices, found numerically by minimizing
the free energy within an elastic string model.  The lengths are in units
of $\lambda$ and the system is isotropic.  
}
\end{figure}

\noindent
can be added at the top of the sample to pin one of the vortices.
We assume that there is a sufficiently small externally applied
magnetic field that there are only two vortex lines within the region of 
interest, an area a few $\lambda$ on a side, where $\lambda$ is the London 
penetration depth of the material.  
The magnetic
dots attract the vortices, so that the top and bottom positions of the
vortices are fixed at the dot locations.
As the AFM is moved along the surface of the sample, it
drags the top of one vortex. 
The force required to drag the vortex line can be directly measured.
When the AFM tip moves in a circular
path around the fixed upper dot,
the two vortices wind together, 
producing an entangled state.  In
Fig.~1(b), we illustrate one possible entangled vortex configuration, obtained
numerically as described below. As the winding angle increases,
the force required to drag the vortex further around increases according
to a form derived below.  If the vortices cut, this force will abruptly
drop.  Thus, using such an experiment, it is possible to directly probe 
whether vortex entanglement can occur.


To estimate the force required to entangle a pair of linelike
vortices, we represent the vortex lines as one-dimensional elastic strings
that cannot cross.  
Such a model may be applicable in the London limit at low fields
$B < 0.2B_{c2}$, although the lack of long-range interactions along
the $z$ axis is a notable limitation 
\cite{Marchetti91b,Sudbo91b,Benetatos9900}.
The vortex configuration is determined by a balance of 
the interaction energy, which drives the strings apart, and
the elastic energy, which pulls the strings together to reduce their
length as they are wound.  Thus, the greater the pitch (defined as the
rate of change of winding angle per vertical distance), the closer
the vortices approach, as they seek the preferred spacing that
minimizes their free energy.  Even if 
they are held radially away from the preferred
spacing, as is shown at the top of Fig. 1(b), they approach the preferred
spacing within the middle of the sample, leading to the roughly constant
spacing in the middle of the sample as shown.
This applies up to a certain pitch; beyond that pitch, we will show that
the vortices are unstable to a collapse.  A similar instability in
a related model was
discussed in Ref.~\cite{Sudbo91}.
The existence within the model of 
a collapse of the entangled state at a singularity when the vortices
approach each other too closely
raises the question of whether the entangled state
can exist in other experimental situations where the vortices are driven
closer than this distance.

{\it Elastic String Model---}
We consider two vortices, at positions $\vec r_1(z),\vec r_2(z)$, with
$\vec r=(x,y)$, giving
the position in the plane as a function of the vertical distance $z$.
As a starting point, we consider the free energy
from the elastic string model \cite{Nelson89}:
\be
\label{f1}
F=
\int\limits_{0}^{L} {\rm d}z\, 
\sum\limits_i
\frac{1}{2}\tilde\epsilon_1 (\partial_z \vec r_i)^2+
2\sum\limits_{i<j}\epsilon_0 K_0(|\vec r_i(z)-\vec r_j(z)|/\lambda),
\ee
with $\epsilon_0=(\Phi_0/4\pi\lambda)^2$, $\tilde \epsilon_1\approx
\sqrt{M_{\perp}/M_{z}}\epsilon_0$.
We fix the boundary conditions on the vortex position at the
top ($z=L$) and the bottom ($z=0$) of the sample, and then minimize
the free energy (\ref{f1}) to find the positions of the vortex in between.
To specify these positions, we need eight real numbers, 
two for each vortex at the
top and another two for each vortex at the bottom.  To simplify, let
us fix two of these coordinates by setting $\vec r_1(0)+\vec r_2(0)=0$; thus,
the origin of the coordinate system is set at the midpoint of the two
vortices at the bottom of the sample.  

We note that if a given pair of functions $\vec r_{1,2}(z)$ minimize the 
free energy for given boundary conditions $\vec r_{1,2}(L)$, then the functions
$\vec r_{1,2}(z)+\vec v z$
also minimize the free energy for a different set of boundary
conditions: $\vec r_{1,2}(L)+\vec v L$.  Thus, it suffices to consider only the 
case in which also $\vec r_1(L)+\vec r_2(L)=0$, as then all other boundary
conditions at the top can also be obtained.  In this case, for
all $z$, $r_1(z)+r_2(z)=0$.
Thus, introduce coordinates with
$r_1(z)=[r(z) \cos(\theta(z)),r(z) \sin(\theta(z))]$ and $r_2(z)=-r_1(z)$.
For the boundary conditions, we fix $r(0),r(L),\theta(0),\theta(L)$.
Then, the extremization of the free energy yields equations of motion:
\be
r^2 \theta_z=J
\ee
\be
\tilde \epsilon_1(\partial_z^2 r-J^2/r^3)
+(2\epsilon_0) K_1(2 r/\lambda)/\lambda=0.
\ee
Exploiting an analogy of
this system to a particle evolving in time, the first equation is the
familiar conservation of angular momentum.

There exist solutions with constant vortex spacing, $r(z)$, and pitch,
$\partial_z \theta(z)$.  To find these, set $\partial_z^2 r=0$ in the
equations of motion to obtain 

\begin{figure}
\begin{center}
\epsfxsize=3.5in
\epsfbox{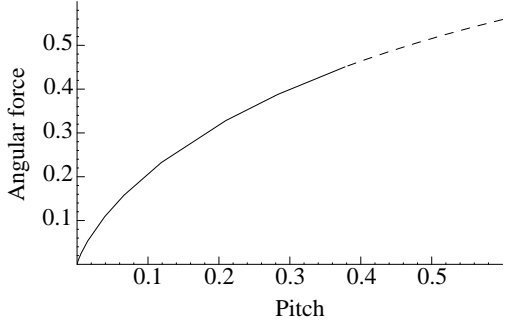}
\end{center}
\caption{Angular force on the top end of each vortex as a function
of pitch $\partial_z \theta$.  
Force is measured in units of $\tilde\epsilon_1$, pitch in
units of $\sqrt{\epsilon_0/\tilde\epsilon_1}/\lambda$.  
The dashed line denotes an unstable vortex configuration
above a pitch of approximately $0.378$,
corresponding to a closest vortex spacing of
$r\approx 1.19\lambda$.}
\end{figure}

\noindent
\be
\label{pitch}
\partial_z \theta=\sqrt{\frac{K_1(2 r/\lambda)}{r \lambda}
\frac{2\epsilon_0}{\tilde \epsilon_1}}.
\ee
This gives the preferred pitch as a function of spacing,
by a balance between elastic energy of one vortex and interaction
energy between two.  For small $r$, this reduces to $\partial_z \theta=
(1/r)\sqrt{\epsilon_0/\tilde \epsilon_1}$.  From this one can show that for
small spacing in the isotropic system ($\epsilon_0=\tilde\epsilon_1$)
the vortices cross at an angle of $\pi/2$,
as expected \cite{Sudbo91}.
While this scenario seems reasonable,
we will find later that there is a major caveat: 
Eq.~(\ref{pitch}) describes a minimum of the free energy 
for large $r$, 
but for
smaller $r$ it describes only an extremum and is unstable to a collapse
of the vortices.

To find the force that would be detected by the AFM tip as the
vortices are wound around one another into an entangled
state,
we first consider the large $r$ case of Eq.~(\ref{pitch})
when
the entwined vortex configuration is stable.  We fix the
total winding angle, $\Delta\theta=\theta(L)-\theta(0)$, 
and the values $r(0),r(L)$.
The force that the 
dragged vortex exerts on the AFM tip can be found by taking a derivative
of the free energy with respect to $r$ and $\theta$.  We find that the 
{\it angular} force on the top end of each vortex is
$\tilde \epsilon_1 r \partial_z \theta(L)$, while the {\it radial} force
on each is $\tilde \epsilon_1 \partial_z r(L)$.  The radial force will
vanish when $r,\partial_z \theta$ obey Eq.~(\ref{pitch}) at the top and
bottom of the sample, giving a solution with constant radius and
pitch $\partial_z \theta=(\theta(L)-\theta(0))/L$.
In this case, we plot the angular force as a function
of pitch in Fig.~2.

{\it Experimental Implications--}
The magnitude of the angular force is large enough to be detected
experimentally.  For example, consider two vortices in a YBCO
sample.  For this material, $\tilde\epsilon_1 \approx 5\epsilon_0$,
where $\epsilon_0 \approx 140$ pN.  Thus, in Fig. 2, a force
of 0.1 $\tilde\epsilon_1$ would correspond to approximately
70 pN, a value within the range of forces detectable with AFM.
In order to convert the pitch $\partial_z\theta$ plotted in Fig. 2
into the total angular displacement $\Delta\theta$
imposed on the dragged vortex, we must know the thickness $L$ of the
sample in the $z$ direction.  Assuming $L=1 \mu$m
gives $\Delta\theta=L \partial_z\theta=0.91\pi$
for $\partial_z\theta=0.2$.
The spacing of the pins at the bottom of the sample does not affect
the force measured at the top of the sample unless it is of order the
sample thickness or wider, due to the fact that the 
vortex spacing in the middle of
the sample depends only on the pitch.

If the radial force is non-vanishing, then
$r(z)$ is not a constant function: by applying a radial force
at the top we drive the vortices away from the preferred spacing.
If $r(L)$ is too
large for the pitch, then $r(z)$ will decrease in the middle of the sample,
while if $r(L)$ is too small, $r(z)$ will increase in the middle of the
sample.  Solutions to the equations of motion can be found 
numerically\cite{shoot}, as in Fig.~1.
For large enough $L$, one finds that in the middle of the
sample the $r,\partial_z \theta$ are given with good accuracy by 
Eq.~(\ref{pitch}), for large enough $r$ when the equation describes
a minimum.

We next consider under what conditions the twisted vortex configuration
becomes unstable.  This occurs when Eq.~(\ref{pitch}) no longer describes
a minimum.
We find by differentiating the free energy twice that for stability
at fixed $J$, we need 
\be
\label{stability}
\tilde\epsilon_1\frac{3 J^2}{r^4}-
\epsilon_0\frac{2}{\lambda^2}(K_0(2 r/\lambda)+K_2(2 r/\lambda))<0.
\ee
When Eq.~(\ref{stability}) vanishes, at $r\approx 1.19 \lambda$,
the system is marginal, and for smaller $r$, the system is unstable to
perturbations.
In this case, the repulsion between the vortices is unable to overcome
the elastic energy, and the two vortices will be driven together, meeting
at a singularity for some $z$ with $r(z)=0$.  
Only a finite energy cost is paid to have
the two vortices meet at a point, while an arbitrary amount of winding
can be accomplished at that point with no free energy cost.
Within an elastic string model,
it can be shown that for any vortex-vortex interaction energy which
diverges less strongly than $1/r^2$ there is an instability of this nature
for sufficiently small $r$.
%
At the marginal $r\approx 1.19\lambda$, the angular force on each vortex
is equal to $(0.45...) \sqrt{\tilde\epsilon_1\epsilon_0}$.  Thus,
this is the largest angular force possible in this experimental setup,
while the minimum stable vortex spacing is $2 r\approx 2.39 \lambda$.

In more highly anisotropic materials such as BSCCO, the elastic string
model for the vortex lines is expected to break down due
to the possibility of decoupling of vortex pancakes on adjacent
layers.  Consider a magnetic AFM tip bound to a pancake at the top of
a vortex line in BSCCO.
If this pancake decouples from
the rest of the 
vortex line below, then the angular force as a 
function of angular displacement will not follow the form 
shown in Fig 2.
but will instead be periodic in the winding angle,
due to the interaction with the remaining stack of
pancake vortices left behind.   
This is illustrated by a simulation of a pancake vortex system 
\cite{bscco} shown
in Fig. 3.  The force from a single pancake
is extremely 

\begin{figure}
\begin{center}
\epsfxsize=3.5in
\epsfbox{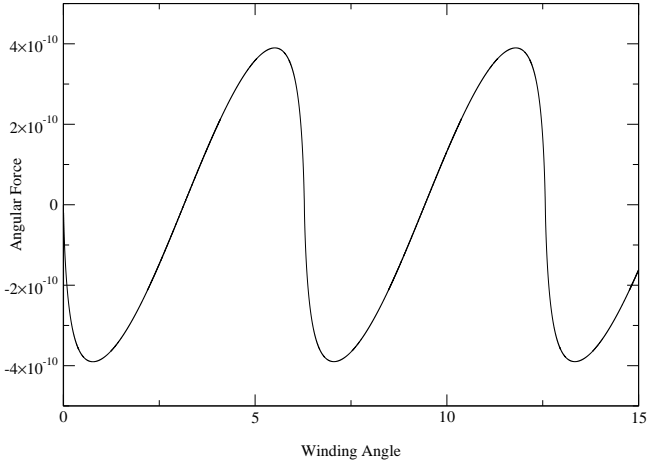}
\end{center}
\caption{Angular force on a decoupled pancake vortex in units of 
$\tilde\epsilon_1$ as a function of
total winding angle $\Delta\theta$, obtained from
a simulation of pancake vortices.}
\end{figure}

\noindent
small, of order $10^{-20}$N; thus 
it is natural to ask how to experimentally determine
whether one pancake is attached to the AFM tip, or whether
the entire vortex stack has detached from the AFM tip so that the tip
is moving freely over the sample.  The
presence of a vortex under the AFM tip can be detected 
by means of a local density of states measurement performed by
temporarily changing the mode of 
operation of the tip to a tunneling probe.  

{\it Discussion---}
The local experimental probe that we propose can also be used to 
explore numerous other properties of the vortex system besides vortex
entanglement.  For example, in a geometry containing only one magnetic
pin and one vortex line, the AFM tip can be used to measure the vortex
line tension directly.  If the line tension is known, the tip could
be used to tear a vortex away from a (weaker) individual pin, such
as a grain boundary, and the pinning force could be measured.  By 
applying a transport current to the sample, the Lorentz force can
be determined directly.  Local rheology measurements are also possible
in the vortex lattice state; for example, the local elastic constants
can be probed by moving a single vortex back and forth around its
lattice equilibrium position.  The temperature dependence of both
the elastic constants and the pinning energy could also be probed.

{\it Conclusion---}
We have proposed an experimental setup for constructing 
and probing entangled states
of superconducting vortices, and shown that the forces 
associated with vortex entanglement are experimentally
measurable.
This kind of experimental setup can be generalized to other types of vortices,
as in fluid turbulence.  
Within the elastic string model, we find that the entangled state
is only stable up to a maximum pitch or minimum vortex spacing.
The instabilities we have found raise the question of whether the entangled
state can exist with a high density of vortices.  To answer this
question and to compare pancake and elastic string models, 
an experimental test
of our proposal is desirable.

{\it Acknowledgments---}
We thank C. Reichhardt for helpful discussions.
This work was supported by DOE grant 
W-7405-ENG-36.


\begin{thebibliography}{99}
\bibitem{shoot} To find such solutions, one can use a shooting technique.
It is useful also to exploit the conservation law:
$\tilde\epsilon_1 [J^2/r^2+(\partial_z r)^2] -(2 \epsilon_0)
K_0(2 r/\lambda)={\rm const.}$
\end{thebibliography}
\end{document}